# Instructor-Created Custom GPTs as Pedagogical Partners Fostering Immersion in Online Higher Education: Two Case Studies

Dennis Beck[1][0000-0003-1771-3237] and Leonel Morgado[2][0000-0001-5517-644X]

[1] University of Arkansas, Fayetteville, USA
[2] Universidade Aberta & INESC TEC, Lisbon & Porto, Portugal
debeck@uark.edu

**Abstract.** As online higher education expands, sustaining student engagement remains a critical challenge. This paper approaches immersive learning by investigating how custom GPTs foster immersion (as a state of deep mental involvement) for students and instructors. While large language models (LLMs) offer potential for enhancing feedback, little research has examined instructor-created custom GPTs designed to align with specific pedagogical goals. This paper addresses this gap, employing the Immersive Learning Cube framework, which conceptualizes immersion through three dimensions: system (envelopment by the environment), narrative (meaningful context), and agency (commitment to meaning-making). Through a qualitative analysis of two distinct case studies, an accelerated graduate grant writing course in the US and an undergraduate software engineering course in Portugal, we analyze course-embedded artifacts to map how custom GPTs influence these immersion dimensions. In the grant writing course, the custom GPT functioned as a feedback partner, fostering system immersion through its immediacy, narrative immersion by reinforcing the proposal's evolving story, and agency immersion by empowering students to negotiate feedback and take ownership of revisions. In the software engineering course, a diegetically-framed custom GPT acted as a metacognitive tutor, enhancing system immersion via its permanent availability, narrative immersion through its role-play function and agency immersion by scaffolding students' self- and co-regulated learning. Our findings demonstrate that thoughtfully integrated custom GPTs can act as powerful pedagogical partners that leverage all three dimensions of immersion. Rather than replacing human instructors, they can amplify immediacy, coherence, and learner autonomy, creating more engaging and immersive online learning environments.

## 1 Introduction

Online higher education continues to expand globally, offering flexibility and access for diverse learners while introducing new pedagogical challenges. Instructors must sustain engagement, provide clear communication, and deliver timely, high-quality feedback, all of which are intensified in active-learning, student-centered courses and, as in our first case study, further compounded by an accelerated format: compressed timelines leave limited opportunity for iterative feedback and reflection. In such contexts, fostering immersion may be a useful scaffold for maintaining meaningful learning experiences. We approach this by positioning instructor-created custom GPTs as an AI-powered immersive learning technology: GPTs as pedagogical partners. Here, we use the term "pedagogical partner" in a functional, non-anthropomorphic sense. We see the learning activities as a form of distributed cognition [1]: a cognitive ecosystem [2] distributed across learners, instructors, and mediating artifacts; in our cases, the custom GPTs participate in that ecosystem by sustaining feedback and interaction within the course activities. In this view, tools are not passive elements but rather exert agency within the ecosystem; instructor-created GPTs can therefore be analyzed as "partner" entities within the designed activity system, alongside human participants and other mediating artifacts (including non-intelligent tools).

This positioning is done by interpreting custom GPTs under the lens of Immersion. Agrawal and his colleagues define Immersion as a state of deep mental involvement that "causes a shift in attentional focus and may lead to a reduced awareness of the physical environment" [3]. Nilsson et al. [4] had previously demonstrated that this phenomenon derived from three conceptual dimensions, which we later proposed to be named system, narrative, and agency [5].

These dimensions provide a useful structure for examining how different learning environments, digital or otherwise, can draw learners into deep attentional engagement. The first is system immersion, the technological system's ability to envelop the participant (e.g., its responsiveness and fidelity), and this includes perspectives on its functionality, such as the timeliness of feedback. The second dimension, narrative immersion, results from contextual meaning (e.g. spatial and temporal elements in a story, emotional engagement with outcomes or elements). In online courses, especially those delivered asynchronously, students often struggle to connect isolated tasks into a cohesive whole [6, 7]. Feedback, whether human or AI-generated, may serve as narrative cues that provide context and reinforce coherence across learning activities. The third dimension, agency immersion emerges from a commitment of the participant to meaning-making (e.g., goal-directed activities, the learner's sense of control over outcomes) [4]. This dimension of immersion depends on the balance between guidance and autonomy. Instructors who structure opportunities for decision-making, collaboration, and self-regulation create conditions for students to experience agency in their learning.

Together, these three dimensions offer a useful framework for examining how technology-mediated feedback systems may promote or constrain engagement in online learning environments. As Agrawal et al. [3] note, immersion is a psychological state that arises from the interaction of these design dimensions rather than from any single element in isolation. Understanding how specific instructional technologies contribute to this interplay is therefore critical for developing effective digital learning environments.

The emergence of large language models (LLMs) such as ChatGPT has renewed scholarly interest in how AI might influence these immersive processes. Existing research highlights LLMs' capacity to provide rapid, adaptive feedback on writing mechanics, structure, and clarity, while also noting limitations such as inaccuracies, overgeneralization, and potential over-reliance by students [8, 9]. When implemented thoughtfully, such tools may enhance system immersion through immediacy and agency immersion through iterative engagement with feedback [10].

Despite growing attention to generative AI in education, little research examines custom GPTs, versions of language models configured by instructors to reflect course expectations, disciplinary conventions, and pedagogical goals (e.g. ChatGPT's MyGPTs, Claude's Skills). While such large language models (LLMs) introduce new opportunities, most studies focus on general-purpose tools like ChatGPT. We address this gap through two case studies of instructor-created custom GPTs in online higher education.

To interpret the pedagogical and experiential aspects of these instructor-created GPTs, this paper draws upon the Immersive Learning Cube framework [11] (ILC), which situates learning experiences within Nilsson et al.'s cube of Immersion dimensions. Nilsson et al. imagined a cube where each dimension, rather than being x, y, z, instead represented each of the three conceptual dimensions of immersion from 0 (no such dimension) to 1 (full account of that dimension). In a previous work in 2020, recently updated [11], we conceived that each of these cube dimensions could be interpreted as percentages of how much a given learning environment was leveraging each dimension of immersion or depended on it. In both the previous work and its update, we placed upon it the identified uses of immersive environments, hence situating them in the three-dimensional conceptual cube. This framework provides a structured method for describing and analyzing immersive learning experiences, making diverse cases comparable across contexts.

This paper applies the ILC to two case studies of custom GPTs used in online higher education. The first, from the United States, examines a GPT employed in an accelerated graduate course on grant writing. The second, from Portugal, explores the use of a custom GPT in an undergraduate informatics course. Though the disciplinary contexts differ, both share the goal of leveraging GPT as a pedagogical partner that enhances feedback, engagement, and immersion.

By framing these cases through the ILC, we analyze how instructor-created GPTs influence the dimensions of immersion in online education. Specifically, we ask the following research question:

*RQ: How do custom GPTs foster immersion for students and instructors in online higher education?*

## 2 Methods overview

We conducted two autonomous case studies in our roles as instructors of the online courses. This study adopts a practitioner-inquiry stance, which relied only on professional practice elements produced in the normal conduct of teaching (forum posts, instructor feedback, and student reflections). No additional intervention or data collection beyond routine instruction occurred. Our approach was to conduct a qualitative content analysis guided by the ILC framework. In the results sections we report what the artifacts directly show; in the analysis & discussion sections we interpret that through the ILC to propose design inferences (plausible mechanisms). Students' immersion is inferred from student-authored elements (post excerpts), whereas instructor immersion is inferred from instructor-authored elements and documented shifts in instructional activity (e.g., giving feedback, narrative choices, interventions). For each case we acted as follows.

We analyzed course-embedded artifacts appropriate to each setting: (Case 1) GPT feedback excerpts, instructor comments, peer-review rubrics, and student reflections; (Case 2) 87 public Moodle forum posts that explicitly referenced "Catmming", a custom GPT focused on providing self-regulation and metacognitive support (described in section 4.1), plus instructor replies and selected team artifacts mentioned in those posts. We included materials created during ordinary teaching; we excluded materials unrelated to GPT use or immersion and any content containing personal/sensitive data. The primary unit of analysis was a post or feedback exchange between AI, students, and/or instructors; where relevant we considered short sequences as a single episode. Syntheses of the prompt structures for each custom GPT are provided in each case's sections.

Each of the researchers independently read through each data for indicative cues of immersion as defined by the ILC framework. After identifying text segments that explicitly referenced GPT use or showed GPT-mediated interaction (e.g., pasted outputs, reported conversations, instructor comments), we classified it with one or more provisional ILC codes (system/narrative/agency) using the ILC definitions. For reporting, we selected illustrative excerpts that were representative of recurring patterns and could be safely presented without identifying the author; where available, we retained pseudonymized IDs and timestamps. Disagreements were discussed until consensus to convergence and reflected in the criteria for classification as system, narrative, or agency, detailed below for each case; we did not establish inter-coder reliability because our goal was design inference.

We then used GPT-5.2 Thinking as a co-intelligent partner to: (i) draft candidate indicator lists from our memos (signals and quotes), (ii) propose tentative classifications for those indicators, and (iii) debate in the role of questioner (GPT) and witness/participant (researchers/instructors). Human authors made all inclusion/exclusion decisions, verified all quotes, and rejected model suggestions when inconsistent with the artifacts. The model had no access to private student data, logs were pseudonymized with "Sn" for each student and "Instructor" for the instructor. Private GPT chats were not accessed and are not accessible to instructors. Quotes were de-identified and checked to avoid personal/sensitive information, following data minimization and GDPR principles.

Findings are grounded in ILC-guided analysis of contextual evidence. Case 1 includes student self-reports and curated feedback exchanges; Case 2 relies on forum traces and cannot link private custom GPT use to outcomes. Evidence may be biased toward engaged participants who posted. We do not claim causal effects or general prevalence; rather, we articulate mechanisms by which custom GPTs plausibly foster immersion in online courses.

## 3 Case Study #1: A custom GPT as a feedback partner in an accelerated online graduate course on grant writing

### 3.1 Introduction

Grant writing is an important professional skill across multiple contexts. Success requires aligning needs and problem statements, goals, methods, and budgets with funder priorities in a persuasive, evidence-based narrative. Teaching students to master this skill is complex, particularly in online accelerated formats where time constraints heighten the demand for rapid feedback and revision.

In Summer 2025, an instructor piloted the integration of a custom GPT model into *An Introduction to Grant Writing*, a fully online, five-week, three-credit course at University of Arkansas (USA). Students worked in groups to produce authentic proposals submitted to external funders. The GPT model, trained to provide structured grant writing feedback, alternated with the instructor's own feedback throughout the drafting process. This collaboration aimed to address two challenges: the difficulty of providing immediate, detailed feedback in accelerated settings, and the need for students to learn to interpret and apply feedback from multiple sources.

This course also provided a unique opportunity to examine how such human–AI collaboration contributes to learner immersion in online higher education. The accelerated, online nature of this course required strategies that could promote all three forms of immersion simultaneously. A custom GPT model functioned as an ever-present feedback partner intentionally structured to sustain student engagement across these immersive dimensions.

This study addresses the overall question of this paper by asking:

*How does integrating a custom GPT as a feedback partner foster students' and instructors' immersion in an accelerated online graduate course on grant writing?*

We address this by mapping artifact-based evidence to system, narrative, and agency dimensions of immersion and by interpreting the GPT's role in those, relative to instructor orchestration and course design.

### 3.2 Background and Literature Review for the case

Grant writing instruction has grown in importance as institutions increasingly rely on external funding to support innovation and research. Yet prior studies highlight the difficulty of teaching proposal writing because it requires both technical precision and rhetorical skill [12]. Collaborative, project-based approaches are common, as grant writing in professional contexts is rarely an individual task [13]. However, these approaches demand substantial instructor guidance to ensure alignment with funder criteria and coherence across proposal sections, challenges that can be intensified in online and accelerated learning environments.

Feedback has long been recognized as one of the most powerful influences on student learning [14]. In writing instruction, feedback not only addresses surface-level mechanics but also supports students' development of argumentation, organization, and critical thinking [15]. Timely feedback, in particular, enhances both the quality of revisions and student motivation [16]. However, in accelerated online courses, compressed timelines limit instructors' capacity to provide frequent, individualized responses, often reducing opportunities for formative dialogue.

### 3.3 Methods

Grant Writing in Educational Technology was delivered in a fully online, five-week summer term. Participants were graduate students in a College of Education at a major university in the U.S.A., working in teams of 4–5. Each team selected a grant topic aligned with their professional or disciplinary interests. Proposals were required to be authentic submissions to real funders, located via the proprietary GrantForward database[1]. Groups collaborated asynchronously and synchronously using Blackboard Ultra, shared cloud documents, and Zoom.

The course followed a scaffolded sequence: brainstorm topic, needs statement, goals and objectives, methods, evaluation, budget, and summary. Each section was drafted, revised, and resubmitted for feedback, resulting in multiple proposal sections being in progress simultaneously. Students submitted drafts of these sections up to three times per week.

The instructor uploaded drafts to the custom GPT helper, which was trained to provide detailed formative feedback aligned to the course's grant writing rubric. GPT feedback was delivered at the group level as part of the drafting workflow and discussed within teams alongside instructor and peer input. The GPT was configured with system-level rules defining it as a grant-writing coach that (i) leverages the instructor's uploaded evaluation records for consistency, (ii) avoids deferred/async responses, (iii) uses web browsing only for time-sensitive verification with citations, and (iv) returns structured outputs (e.g., critiques + rewritten text + measurable objectives) tailored to the user's request.

GPT feedback was reviewed for accuracy before being released to student groups. Instructor feedback followed within 24-48 hours, addressing higher-order issues such as argument coherence, funder alignment, and rhetorical persuasiveness. This alternating feedback structure (immediate AI guidance followed by deeper instructor critique) was designed to balance system immersion (through immediacy and responsiveness) with agency immersion (through reflective interpretation and revision).

During week four, groups constructed funder-specific rubrics alongside a near-final draft of their proposal. These rubrics, grounded in real funder criteria, were used by two peer groups to evaluate one another's work. Peer reviews were followed by live Zoom proposal presentations and discussions, further extending narrative immersion through authentic professional simulation.

This analysis draws on multiple forms of qualitative evidence documenting the feedback and revision process, including:

1. GPT-generated feedback records from all major proposal sections;
2. Instructor comments on student work across all feedback rounds;
3. Observational notes from student interactions during peer review and proposal presentations; and

---
[1] https://www.grantforward.com

4. Student discussion transcripts and course reflections, including the final class discussion in week five, where students were asked to reflect on their learning process and the role of the custom GPT in supporting their work.

All data sources were analyzed using the ILC [11] as a guiding framework, as specified earlier. Specifically, course materials, feedback exchanges, and discussion reflections were analyzed for indicators of:

- System immersion: engagement with the online course environment and immediacy of feedback cycles by the custom GPT;
- Narrative immersion: the context of a real-world grant application, emotional investment, coherence, and continuity across proposal development;
- Agency immersion: a topic that learners' decide as meaningful for them, learner autonomy, reflection, and decision-making in response to feedback.

Emergent themes were triangulated across these dimensions to examine how the custom GPT contributed to (or constrained) immersion in each category. This immersion-centered lens allowed for holistic interpretation of both student and instructor experiences, situating the findings within broader theoretical understandings of engagement in AI-supported online learning.

## 3.4    Results

**System Immersion: Immediacy and Interaction.** System immersion was primarily generated through the immediacy and consistency of GPT feedback. An example of this was when students were asked to submit their grant proposal goals and objectives. In one case, initial GPT feedback suggested that two of their objectives were not measurable, and suggested specific ways they could revise those objectives (Uark, Grant writing team one, June 3, 2025). Another team's objectives were vague on the deadline for these objectives, and suggested appropriate deadlines be added (Uark, Grant writing team three, June 3, 2025). Students from Grant Writing Team One described the experience as "like having another instructor on call," emphasizing how rapid feedback maintained their focus and momentum (Uark, Grant writing team one, June 28, 2025). Another student reflected that "the GPT kept us working. It was impossible to procrastinate because we always had something to fix right away." (Uark, Grant writing team three, June 3, 2025)

The GPT's responsiveness reduced downtime between drafts and sustained engagement in a course where time was the scarcest resource. Another student wrote, "Five weeks felt intense, but the constant feedback kept me immersed. There wasn't time to drift off, we were always working on something." This immediacy created a sense of conversational flow within the system, aligning with Nilsson's description of system immersion as responsiveness that reinforces a learner's presence in the activity.

Instructor reflections echoed this benefit: "ChatGPT's feedback gave me the ability to focus more on the bigger picture, where I excel anyway." The AI thus extended system capacity, enabling continuous feedback cycles that deepened engagement without diminishing human oversight.

**Narrative Immersion: The Story of a Proposal.** As students progressed through successive revisions, their proposals evolved into coherent narratives, a shift that reflected growing narrative immersion. Initially, many students conceived of sections as discrete tasks ("we need X"), but as GPT feedback repeatedly connected needs, goals, and methods, students began to perceive their proposals as unified arguments. For example, one grant writing team submitted a proposal that received the following feedback: "Proposal reads as separate sections written by separate authors. Consider having the same person edit the entire proposal to ensure that sections transition well into the next one." (Uark, GrantGPT, June 17, 2025). On the evaluation of another proposal, GrantGPT encouraged students to "create a concept map which connects your objective to your methods to your results to your evaluations." (Uark, GrantGPT, June 19, 2025). A student from Team Three student noted in response to this feedback, "I used to think grant writing was just filling in boxes, but the way GPT and [the instructor] (…) kept connecting our sections made me see how it's all part of one story." (Uark, Grant writing team three, June 28, 2025)

This sense of narrative progression fostered emotional investment. Students described feeling "in the trenches" with GPT, and another said, "It was frustrating at times, but kind of addictive once we started seeing improvement every week." The emotional dimension of narrative immersion was therefore reinforced by the sense of iterative progress and shared authorship between humans and AI.

**Agency Immersion: Collaboration, Control, and Learning Autonomy.** Agency immersion emerged through students' increasing ability to create meaning by directing their learning and negotiating multiple sources of feedback. Working in groups, they first had to choose their own topic for the grant proposal, as well as the funder to whom they were going to apply. This gave the students a high level of investment in the success and completion of the grant proposal, because it meant so much to them. As they interacted with the custom GPT, student groups also balanced the GPT's immediate suggestions, instructor critique, and peer perspectives. For example, Team Four received GPT feedback on their methods that suggested inclusion of some very important, and other non relevant details (Uark, GrantGPT, June 7, 2025). In response, a student from that group explained, "We had to decide what to accept or reject: GPT gave options, but we had to choose what made sense." This negotiation process cultivated ownership and self-efficacy. Another student said, "It felt like a conversation, not a correction." (Uark, Grant writing team four, June 28, 2025).

The instructor intentionally framed GPT feedback as a starting point for debate, not a final authority. Students from Team Two reported feeling empowered by this framing: "GPT told us what to fix right away; [the instructor] (...) explained why it mattered." (Uark, Grant writing team two, June 28, 2025). Over time, they learned to anticipate the GPT's prompts, demonstrating meta-cognitive engagement with the writing process. As another student from the same group wrote, "By the end, I knew what GPT would probably say. I started fixing those things before even running it." (Uark, Grant writing team two, June 28, 2025)

Such comments reflect the development of agency immersion: students felt in control of both the process and the learning outcomes, despite the accelerated pace. The iterative cycle of feedback, reflection, and revision became a space of active inquiry rather than passive correction.

### 3.5 Analysis via the Immersive Learning Cube

To further interpret how immersion manifested in this course, the ILC [11] was applied to categorize design elements and learner experiences across the three dimensions of immersion (Table 1).

Table 1. Mapping design elements and observed indicators to Case#1 immersion dimensions

| Immersion Dimension | Course Design Elements | Observed Indicators of Immersion | Illustrative Evidence |
|---|---|---|---|
| **System Immersion** | Custom GPT integrated into workflow; immediate feedback following each submission; Instructor review provided 1-2 days after AI response. | Continuous engagement with course system; perception of "always-on" feedback loop; reduced downtime between drafts. | "The GPT kept us working. It was impossible to procrastinate because we always had something to fix right away." |
| **Narrative Immersion** | Authentic, real-world proposal development; iterative storytelling through linked proposal sections; feedback framed as continuing a dialogue. | Emotional investment and sense of shared authorship; perception of the proposal as an evolving story. | "It felt like the GPT was in the trenches with us… it was frustrating at times, but kind of addictive once we started seeing improvement every week." |
| **Agency Immersion** | Group collaboration; explicit framing of GPT as a tool, not authority; instructor prompts for critical decision-making. | Increased learner autonomy and confidence in managing multiple feedback sources; negotiation and ownership of revisions. | "GPT told us what to fix right away; [the instructor] explained why it mattered… by the end, I started fixing things before even running it." |

This analysis shows that system immersion was most directly enhanced by GPT's immediacy, narrative immersion by the coherence and emotional momentum of proposal development, and agency immersion by students' growing control and reflective decision-making. The layered feedback structure thus activated all three dimensions simultaneously, creating conditions for sustained immersion in an accelerated online environment.

## 3.6 Discussion

Framed through the three dimensions of Immersion, the integration of a custom GPT in this case significantly fostered immersion in this accelerated online course. System immersion was achieved through immediacy and environmental interactivity, the GPT's constant responsiveness reduced cognitive friction and sustained attention. Narrative immersion arose as feedback cycles created a continuous storyline of improvement, with students invested in their evolving, real-life proposals. Agency immersion developed as students exercised control over revisions, negotiated inputs, and internalized both AI and instructor feedback.

These findings support Agrawal et al.'s [3] assertion that immersion results also from attentional and emotional engagement rather than technological novelty alone. The GPT did not create immersion by itself; rather, it extended and stabilized the immersive conditions already cultivated through sound pedagogy, scaffolded assignments, authentic tasks, and consistent instructor presence.

From the instructor's perspective, the deliberate positioning of GPT as a feedback partner (not co-instructor) preserved pedagogical authority and human connection. This was crucial for sustaining trust and preventing over-reliance, risks noted in prior studies of AI feedback [9]. By mediating between human and AI feedback, the instructor maintained narrative coherence and reinforced the reflective dimension of agency immersion.

## 3.7 Conclusion

Integrating a custom GPT into an accelerated graduate grant writing course fostered immersive learning across system, narrative, and agency dimensions by sustaining engagement, rapid progress, and a sense of ownership over their work. The instructor benefited from redistributed workload and greater capacity to focus on higher-order critique.

The findings suggest that AI can foster immersion not necessarily by replacing human feedback but by amplifying immediacy, coherence, and learner autonomy. Future iterations will further explore structured prompts that encourage students to critically evaluate AI feedback, ensuring that immersion remains both reflective and authentic.

## 4 Case Study #2: Metacognition, and Self- and Co-Regulation tutoring in software engineering education

### 4.1 Introduction

Team-based software engineering is a core professional competency. Competent practice requires design discipline and quality attributes (cohesion, low coupling, separation of concerns) together with systematic testing, robust error handling, and clear interface–implementation boundaries. Teaching the shift from craft-like solo coding to collaborative engineering is challenging in any setting because the value of these qualities emerges under real-world change and maintenance, conditions that are hard to simulate within mid-program undergraduate courses.

The course *Software Development Laboratory* at Universidade Aberta (Portugal) is fully online, lasting a full semester of the Informatics Engineering baccalaureate (2nd year, 2nd semester), and delivered in the Moodle LMS under the university's Virtual Pedagogical Model, which emphasizes student-centered learning, flexible asynchronous interaction, and diversified forms of communication [17]. The course, with ~200 students (ages approx. 25-55, working students), uses the e-SimProgramming didactic approach, which employs a text-only narrative immersion role-play in which the instructor acts as CEO, the human tutor as CTO, and students as trainees in a simulated software development company. It focuses on developing skills for self- and co-regulation of learning (SCRL), including metacognition [18], and providing formative assessment; students are expected to assume active responsibility for their project-based learning while working in collaborative teams [19]. For scaling the approach, providing on-demand feedback including outside business hours (a common need of online students), the instructors introduced Catmming, a custom GPT positioned diegetically as the company's "intelligent automatic assistant." Catmming is present on the course home page with the label "Metacognitive support," on par with the instructors, and students were encouraged to use it privately for planning, monitoring, and reflecting on their work.

This study addresses the overall question of this paper by asking:

*How does integrating a custom GPT as a self-regulation and metacognitive tutor foster students' and instructors' immersion in an online undergraduate course on software engineering?*

### 4.2 Background and Literature Review for the case

In fully online education, students must plan, monitor, and adapt their learning with fewer incidental cues and less temporal synchrony than on campus, making self-regulated learning (goal-setting, monitoring, strategy adjustment) foundational to progress and persistence [20], a reality that is also found in online software engineering courses [18]. Because code must evolve, be handed over, and integrate into larger systems, the payoff of design discipline and quality attributes becomes salient under change and maintenance; helping students notice and reason about that payoff is the role of metacognition in programming and software engineering [21].

In team settings, regulation is also social: roles, shared artifacts, and feedback rituals enable co-regulation, shaping peers' goals and strategies; recent work in software development education documents how collective regulation and team cognition relate to engagement and performance [22]. Large online cohorts, however, create a practicality problem: timely, individualized dialogue scales poorly across time zones, and evening/weekend study. Evidence from AI-enabled settings suggests a partial remedy: when configured to local norms and prompts, LLM-based assistants can provide formative feedback and metacognitive scaffolds, improving strategy use and engagement [23].

### 4.3 Methods

Software Development Laboratory ran online from March to June 2025 in Moodle. Following the e-SimProgramming didactic approach and learning objectives and content outlined in this case's introduction, students worked in role-defined teams of five (1 Leader, 3 Developers, 1 Verifier) on software-development projects. Each team defined a project aligned with its interests under the fictional company scope: "*The company wants to explore the potential of several APIs for new products or features. We need you to try them out and propose a small demo application so we can discuss that.*" Teams coordinated via Discord, submitted deliverables in Moodle and maintained code on GitHub. All course contents and interactions were framed diegetically within the fictional company, with fictional colleagues and situations.

Across the semester, activities followed a predictable two-week topic rhythm: planning and design tasks, implementation tasks (including testing/error-handling), and metacognitive challenges. The latter ran, throughout, as graded, lightweight quizzes: (a) fortnightly reflections that asked students to monitor progress, difficulties, and next steps, and (b) content-linked prompts that elicited justification or rethinking of design and development choices. These challenges functioned as regular plan/monitor/reflect touchpoints within project work, not as add-on assignments. Continuous assessment accrued portfolio points from these ongoing activities; there were three moments at which the accumulated portfolio was marked, allowing students to iterate until those grading occasions using instructor and peer feedback in open forums. Timing was flexible to accommodate online learners' schedules.

The custom GPT was part of this role-play: Catmming, the company's "intelligent automatic assistant". The Catmming GPT was (i) listed and linked to, on par with the professor and tutor, on the course homepage as "Metacognitive support", (ii) its use was routinely encouraged by the lecturer and resurfaced by students in forum posts, and (iii) the character (not the GPT) was named in static text as a speaking character in the narratives inside the metacognitive quizzes. Students were invited to use the Catmming GPT privately for planning, monitoring, and reflection, with explicit assurances that teaching staff could not access those conversations. These metacognitive structures and their cadence follow the established e-SimProgramming didactic practice for fostering SRL/Co-RL in software engineering courses. This custom GPT was configured with a detailed instructional prompt specifying its role as a metacognitive AI tutor (called Catmming) in a simulated software engineering course, using the European Portuguese variety and grounded in Zimmerman's model of self-regulated learning. Constraints included strict linguistic adherence, role-play integration, and staged clarification before action, to foster learner support within a narrative-based e-learning context. The GPT was used privately by individual students but its uptake became socially visible through forum posts and team decision-making, where excerpts were shared, debated, and contextualized by peers and instructors. This analysis draws on 87 such Moodle forum posts, with multiple forms of qualitative evidence documenting how Catmming was used and taken up, including:

1. Students' written accounts of their Catmming GPT conversations posted in Moodle forums;

2. Student-shared excerpts of Catmming's feedback attached to those posts; and
3. Instructor responses to forum interventions involving Catmming.

These sources let us sample the custom GPT's feedback itself, observe how students involved it and responded, and see how instructor participation was shaped by its presence. Over the term, the custom GPT's "About" page reported "400+" Catmming conversations (no finer breakdown available). The public rating distribution comprised six votes, heavily skewed toward 5-star ratings, with single 4- and 3-star ratings and no 1-2-star ratings; these figures indicate perceived usefulness, not formal evaluations. For interpretation and triangulation, we also drew on our experience running the course during the semester and consulted selected team artifacts (e.g., deliverables and plans) to corroborate forum-reported aspects.

All forum-based data were analyzed using the Immersive Learning Cube [11] as the guiding framework. We looked for indications of:

- System immersion: responsiveness, immediacy, reduction of cognitive friction;
- Narrative immersion: diegetic coherence, storyline continuity, identity within the simulation, and
- Agency immersion: learner choice, regulation, negotiation of alternative solutions.

Because OpenAI does not provide per-session usage logs for the custom GPT, we treated post timestamps as a proxy for outside-hours engagement (evenings after 19:00, early mornings before 08:00, and weekends when marked as Saturday or Sunday). Since private Catmming chats are not made available by OpenAI; their attribution as cause was only made when students explicitly linked outcomes to Catmming in their forum posts. Where appropriate, we triangulated interpretations with selected team artifacts (e.g., plans/deliverables) referenced in the forums.

### 4.4 Results

**System Immersion: responsiveness that extends the learning day.** Evidence from the artifacts suggests Catmming reduced friction and provided just-in-time explanations when students were most active (nights and weekends) with many Catmming-referencing posts stamped late evening (e.g., 23:51; 23:02) or weekend days. Students reported using the assistant for rapid bootstraps, e.g., C++ vs. C# programming language differences; course matter summaries on the Model-View-Controller (MVC) architectural style and input/output mapping, and for crisp analogies that stabilized understanding (e.g., "*It's like an oven alarm*", UAb Sn6, April 8, 2025, 22:28). Catmming's constraints were also surfaced and absorbed into practice (mentions that Catmming's session history starts 'clean'; occasional knowledge base gaps), which the instructor then addressed in-course (e.g., patching the knowledge base), keeping interaction predictable without displacing human oversight, reinforcing a transparent, predictable interaction loop. The instructor also reiterated Catmming's role as metacognitive aid rather than authority ("it's for metacognitive support, not a coordinator"), which we interpret as having helped maintain low friction while setting expectations.

**Narrative Immersion: sustaining the fictional world and role expectations.** The SimProgramming fictional company diegesis framed activity consistently: students acted in role ('leader' or 'spokesperson' mentioned), discussed team rituals, and reposted Catmming content that positioned itself as tuned to the course context. Where students contrasted tools, they typically flagged uncertainty about generic GPTs and anticipated Catmming's better fit because it was course-specific (some noted it was developed by the instructor). For instance: "*I take the opportunity to share an output about MVC that I requested to better understand. (I used the GPT-4o from OpenAI base without additional resources like in Catmming, so \*\*it may contain errors\*\*)*.", UAb Sn16, Sunday, March 16, 2025, 01:01. The instructor mentioned Catmming as a character in a Carnival holiday announcement, for narrative continuity; students did not echo that motif, but the course-world cadence remained coherent through ongoing diegetic messaging and the quizzes where the Catmming character 'speaks' (static text).

**Agency Immersion: negotiating design trade-offs and regulating work.** Students reasoned through alternatives after consulting Catmming, then justified choices to peers: whether/when to introduce the 'interfaces' programming construct, how to bound persistence without "scattering responsibilities," when to use events/delegates vs. interfaces to manage coupling, architectural variants, and the course content alongside other architectural alternatives, like microservices. E.g.: "*This topic of events and delegates has only now started to become less 'strange'/'confusing' for me, (…) I decided to confirm with 'Catmming,' I leave the answer here.*"

(UAb Sn4, Saturday, May 10, 2025, 19:20) The instructor routinely returned ownership ("depends on your project and your intents for it"), modeling contextual trade-off thinking. Catmming's planning prompts (as reposted by students) scaffolded team regulation (leader checklists, task split, reviews). Students also calibrated reliance, explicitly tagging generic GPT outputs as possibly erroneous and seeking confirmation, a pattern consistent with reflective uptake rather than deference.

### 4.5 Analysis via the Immersive Learning Cube

To further interpret how immersion manifested in this course, as in Case #1, the ILC [11] was used to categorize design elements and learner experiences across the three dimensions of immersion (**Table 2**).

Table 2. Mapping design elements and observed indicators to Case#2 immersion dimensions

| Immersion Dimension | Course Design Elements | Observed Indicators of Immersion | Illustrative Evidence |
|---|---|---|---|
| **System Immersion** | Custom GPT (Catmming) linked as "Metacognitive support"; always-available forum workflow; awareness of limits (session reset, knowledge base scope) | Late-night/weekend posting around Catmming; rapid clarifications via student-shared Catmming replies; instructor patches to the knowledge base. | Timestamps like 23:51 / "Saturday"; student quotes of Catmming's oven alarm analogy; instructor note: "Catmming was excellent! … I've added that indication to Catmming." |
| **Narrative Immersion** | SimProgramming diegesis (Boss, CTO, Catmming personae); metacognitive quizzes with the Catmming character; role-based teamwork | Students act/ask in-role (leader/spokesperson); students repost Catmming's claim of course contextualization; contrasts with generic GPTs; instructor calendar cues sustain world rhythm | Student post sharing Catmming's post mentioning "advantage… adapted to the course context"; student caveat "may contain errors" about GPT-4o; instructor Carnival post (no student uptake) as world pacing. |
| **Agency Immersion** | Prompts that invite choice/justification; Catmming framed as support, not authority; peer debate with instructor contextualization | Students weigh trade-offs (interfaces; persistence boundaries; alternative constructs; architecture variants); team self-organization (task split, reviews); explicit validation of GPT outputs | "The criterium must always be a practical one…" (student synthesis after using Catmming); instructor reframing persistence scope ("depends on your project"); leader planning/checklists reposted from Catmming guidance. |

### 4.6 Discussion

Using the ILC lens, Catmming's effects align with three complementary mechanisms. System immersion was supported by out-of-hours availability and just-in-time clarifications that students then reposted to stabilize shared understanding; this accords with evidence that timely guidance sustains progress in online SRL cycles [20] and with immersion as reduced cognitive friction [3]. Narrative immersion benefited from Catmming's diegetic role: feedback and metacognitive prompts arrived "in-universe," helping connect design, implementation, testing, and reflection into one internship storyline, useful against the fragmentation risks noted in online courses [6, 7] and consistent with immersion as an emergent property of coherent design dimensions [11]. Agency immersion emerged as students publicly negotiated Catmming's suggestions, justified trade-offs, and flagged generic-GPT errors: behaviors that map to SRL/Metacognition in programming [21] and to co-regulation via shared artifacts in software teams [22]. Importantly, the instructor's framing of Catmming as aid, not authority aligns with findings that LLMs are most valuable as formative, reflective assistants rather than answer engines [23].

We cannot time-link private GPT usage to specific outcomes; inferences rest on forum evidence and course cadence. Still, triangulated signals across posts and instructor interventions suggest that a custom, diegetic GPT can amplify, rather than replace, pedagogical structures that foster immersion in online software engineering.

### 4.7 Conclusion

In this online software engineering course, the custom, diegetic GPT (Catmming) acted as a metacognitive tutor that amplified (not replaced) the existing pedagogy. It sustained system immersion through always-available, just-in-time clarifications; reinforced narrative immersion by delivering feedback inside the course storyline; and deepened agency immersion as students publicly negotiated its suggestions, articulated trade-offs, and coordinated team action. The instructor's consistent framing of Catmming as aid rather than authority preserved human guidance while extending feedback reach beyond business hours.

Evidence is necessarily limited to forum traces and cannot time-link private GPT use to outcomes, yet triangulated signals indicate meaningful contributions to SRL/Co-RL, momentum, and shared understanding at scale. Practically, this case suggests that custom GPTs can be effective when embedded diegetically, aligned with course cadence, and tasked with metacognitive scaffolding, freeing instructors to focus on higher-order coaching by keeping immersion coherent across system, narrative, and agency.

## 5 Overall Discussion Comparing/Contrasting the Cases and Implications

Across two distinct online contexts, instructor-created custom GPTs served as pedagogical partners that could foster immersion, yet they did so through different mechanisms shaped by course design and by how each GPT was positioned.

**System Immersion.** Immediacy was a designed feature in both settings but manifested differently. In Case 1, learners explicitly reported a near "always-on" feedback rhythm and described momentum gains. In Case 2, we cannot time-link private use of the custom GPT to outcomes; however, posts referencing it often appeared at late-evening/weekend times, and students reposted just-in-time clarifications. Taken together, the cases suggest that availability and fit-for-purpose guidance reduced cognitive friction by aligning to course cadence.

**Narrative Immersion.** Both courses embedded GPT use in meaningful 'storylines', but at different levels of world-building. Case 1 leveraged an authentic professional arc (real grant proposals), where combined human and AI feedback advanced a single evolving argument. Case 2 used a diegetic company simulation; Catmming's character as a GPT and as metacognitive quizzes' 'voice' helped maintain world continuity even when content was technical. The implication is that contextual framing (authentic or fictional) shapes how feedback becomes part of the learner's unfolding story.

**Agency Immersion.** In both cases, instructors framed the GPT as an aid, not an authority, prompting learners to weigh options and justify choices. Case 1 evidence includes students negotiating AI vs. instructor vs. peer feedback on persuasive coherence. Case 2 shows students publicly debating trade-offs after consulting the custom GPT and explicitly caveating generic GPT outputs. Designing GPT interactions as scaffolds rather than as answer engines appears pivotal.

From the instructors' perspectives, GPT integration also reshaped immersive teaching presence. By automating surface-level feedback, GPTs enabled instructors to focus on higher-order concerns (argumentation, coherence, metacognition) thus allowing them to remain engaged at the narrative and agency levels of immersion. The resulting human-AI partnership expanded both the scale and quality of feedback without diminishing instructional authority or human connection. Across cases, instructors reported feeling more able to act as mentors and designers of immersive learning environments rather than as graders or copy editors, as reflected in instructor-side aspects such as feedback orchestration or clarifying expectations.

There are implications for practice. The common design move is to embed the GPT as a scaffold inside the instructional narrative, with the instructor curating coherence and norms. The pairing of design choices and instructor stance yields several actionable heuristics:

- **Embed diegetically and visibly:** Make the GPT's role explicit, with privacy assurances and guardrails.
- **Align to course rhythm:** Alternate rapid AI nudges with human higher-order critique (Case 1) or weave AI with regular reflection prompts (Case 2).
- **Preserve human narrative authority:** Instructor mediation maintains coherence and trust while encouraging student choice.
- **Maintain/patch knowledge bases:** Monitor surfaced gaps and update the GPTs' context to keep interactions predictable and course-specific.

These are design inferences, not generalized effects; they emerged from these two cases.

## 6 Overall Conclusion

Returning to our research question ("How do custom GPTs foster immersion for students and instructors in online higher education?"), these two cases indicate that instructor-created GPTs can foster immersion when they are pedagogically embedded into courses, rather than treated as generic, stand-alone tools. Under the ILC framework, the same approach occurs in both contexts: the GPT is positioned to strengthen the course's system envelopment (timely, low-friction guidance matched to course cadence), narrative coherence (feedback that advances an authentic or diegetic storyline), and learners' agency (scaffolds for planning, monitoring, and principled choice).

For students, this meant sustained momentum, clearer through-lines, and practiced justification of choices. For instructors, it meant reclaimed attention for higher-order critique and narrative stewardship, plus a scalable way to sustain presence across asynchronous schedules. The GPTs amplified (rather than replaced) human teaching presence: they extended reach for routine or formative guidance while freeing instructors to focus on higher-order coaching, sense-making, and affective presence. Evidence constraints differ by case (e.g., direct student testimony in Case 1; forum-trace triangulation in Case 2), so conclusions should be read as design-level claims supported by contextual data rather than universal effects.

Thus, considering prior work on LLMs' ability and deficiencies in providing feedback [8–10], our contribution is to focus specifically on instructor-created, course-embedded custom GPTs and to suggest how their pedagogical integration aligns with system, narrative, and agency dimensions of immersion in online courses. This supports instructional design by treating custom GPTs as course "partners" in a cognitive ecosystem and suggests an implication for immersive learning frameworks: that the ILC can be applied to AI-mediated learning, even outside of virtual or extended reality contexts, by treating custom GPTs as mediating artifacts.

## 7 Future work

Future work will expand this inquiry into a coordinated, multi-course research program examining how custom GPTs can foster immersion across disciplines, cultures, and learning modalities. The next phase of collaboration between University of Arkansas and Universidade Aberta will involve deploying custom GPTs across multiple Educational Technology and Informatics courses, allowing for comparative analysis of how different custom GPT designs influence system, narrative, and agency.

A critical aspect will be to better formalize what "pedagogical partner" means operationally when deploying in online courses the epistemological stance of cognitive ecosystems. While this is the current motivation for the term, this paper does not attempt to specify the distinguishable partnership roles or their boundaries in practice. Future work should therefore clarify GPT partnership roles (e.g., partner-as-feedback-loop participant, partner-as-diegetic actor, partner-as-metacognitive scaffold) together with operational criteria for when and how such partnership functions should be designed and enacted.

The Immersive Learning Cube [11] will guide this phase as both an analytical and design framework, enabling systematic mapping of GPT interactions and immersion indicators across contexts. Through this approach, we aim to develop validated design heuristics and evaluation tools that educators can use to build and assess custom GPTs purposefully aligned with immersive learning principles. However, in this study, immersion is inferred indirectly from textual artifacts rather than measured as an experienced state. Future work should triangulate these artifact-based inferences with additional sources to better characterize the phenomenology of immersion.

Further studies will also explore how sustained exposure to GPT-based scaffolding influences long-term metacognitive growth, self- and co-regulation, and the evolving roles of instructors in AI-mediated environments. Additionally, ethical considerations, such as transparency, authorship, and equitable access, will be investigated as integral aspects of immersive design, ensuring that human judgment and empathy remain central to technology-enhanced education.

Ultimately, this ongoing collaboration seeks to establish a comprehensive framework for AI-mediated immersive learning, bridging educational technology and computer science perspectives. By articulating how custom GPTs foster immersion through system responsiveness, narrative continuity, and learner agency, this work contributes not only to theory but also to the practical shaping of AI-enhanced pedagogy in online higher education.

**Acknowledgments.** We thank the students and teaching teams at University of Arkansas and Universidade Aberta for their engagement. We also acknowledge GPT-5 Thinking as a co-intelligent partner used to structure analyses as described in the methods overview section and used to help draft and edit the text of this paper. All analytic decisions and interpretations are the authors' own.

This work has received funding from the European Union (EUROPEAN EDUCATION AND CULTURE EXECUTIVE AGENCY – EACEA) under grant agreement No 101177241.

**Disclosure of Interests.** The authors have no financial competing interests to declare. During the reported work, Dennis Beck served as instructor of the accelerated graduate course at University of Arkansas and Leonel Morgado served as instructor of the undergraduate Software Development Laboratory at Universidade Aberta; both roles are part of their normal academic duties. The custom GPTs described were created and maintained by the authors solely for instructional purposes; no external funding, equity, payments, or in-kind support were received from OpenAI, GitHub, or other vendors referenced.